\documentclass{aa} 
\usepackage{natbib}
\usepackage{graphicx}
\bibpunct{(}{)}{;}{a}{}{,}
\usepackage{epsf}
\newcommand{\be}{\begin{eqnarray}}
\newcommand{\ee}{\end{eqnarray}}

\usepackage{color}
\definecolor{red}{rgb}{0.7,0,0}
\definecolor{blue}{rgb}{0,0,0.7}

\begin{document}

\title{Relativistic cross sections of mass stripping and tidal 
disruption of a star by a super-massive
rotating black hole}
\author{P. B. Ivanov\inst{1,2}\and
M. A. Chernyakova$^*$ \inst{3,4}}
%\offprints{P. B. Ivanov  \email{pavel@lukash.asc.rssi.ru}}
\institute{Astro Space Centre, P. N. Lebedev Physical Institute,
 4/32 Profsoyuznaya Street,
 Moscow, 117810, Russia\and
Department of Applied Mathematics and Theoretical Physics,
University of Cambridge,
Centre for Mathematical Sciences, 
Wilberforce Road, Cambridge, CB3 0WA, UK\and
INTEGRAL Science Data Centre, Chemin d'\'Ecogia 16, 1290 Versoix, Switzerland
\and
 Geneva Observatory, 51 ch. des Maillettes, CH-1290 Sauverny, Switzerland 
}
\offprints{P.Ivanov@damtp.cam.ac.uk\\ $^*$M.A.Chernyakova is on leave from Astro Space Center 
of the P.N.~Lebedev Physical Institute,  Moscow, Russia}

\authorrunning{P. B. Ivanov\&M.A.~Chernyakova}
\titlerunning{Relativistic cross sections of tidal disruption}

\date{Received ..., 2005/Accepted ..., 2005}

  \abstract
   {We consider the problem of tidal disruption of
    a star by a super-massive rotating black hole.
    
    Using a numerically
    fast Lagrangian model of a tidally disrupted star developed  
    in our previous works, 
    we survey the parameter space of the
    problem and find regions  where the total
    disruption of the star or a partial mass loss from the star takes
    place as a result of fly-by around the black hole.

    Our treatment is based on General Relativity, and we consider
    a range of black hole masses where the tidal disruption 
    competes with the relativistic effect of direct capture of 
    stars by the black hole. We model the star
    as a full polytrope with $n=1.5$ with the solar mass and radius. 
    We show that our results can also be used  to obtain 
    the amount of mass lost by stars with different 
    stellar masses and radii.

   We find that the results can be conveniently represented on
   the plane of specific orbital angular momenta of the star 
   $(j_{\theta}, j_{\phi})$.  We calculate the contours of 
   a given mass loss of the star on this plane,
   for a given black hole mass $M$, rotational parameter
   $a$ and inclination of the trajectory of the star with respect to the
   black hole equatorial plane. In the following such contours are 
   referred to as the tidal cross sections.

   It is shown that the tidal cross sections can be approximated as
   circles symmetric above the axis $j_{\phi}=0$, and
   shifted with respect to the origin of the coordinates in the
   direction of negative $j_{\theta}$. The radii and shifts of these
   circles are obtained numerically for the black hole masses in the
   range $5\cdot 10^{5}M_{\odot}-10^{9}M_{\odot}$ and different values of
   $a$. It is shown that when $a=0$  tidal disruption takes place for 
   $M < 5\cdot 10^{7}M_{\odot}$ and when $a\approx 1$
    tidal disruption is possible for $M < 10^{9}M_{\odot}$.  }
   
\maketitle
\authorrunning{P. B. Ivanov, M.A.~Chernyakova}
\titlerunning{Relativistic cross sections of tidal disruption}

%\maketitle

\section{Introduction}

Tidal disruption of stars by a supermassive
black hole may be very important in galactic centres. It
may provide enough gas to fuel central
engines of AGNs, e.g. Hills (1975). It also may account for non-stationary flashes of
radiation observed in certain non-active galaxies, see Rees (1988) for theoretical discussion 
and  Komossa et al. (2004) for observations. 
  
Tidal disruption takes place when the periastron of the stellar orbit is smaller than
the Roche tidal radius
\be  r_{T}=\sqrt[3]{{M\over m}}R_{st}\approx 1.5\cdot 10^{13}\sqrt[3]{{M_{7}\over m_{*}}}R_{*}cm 
\label{eq i1}, \ee 
where $M$ is the black hole mass and $m$ is the mass of star, $R_{st}$ is the stellar radius,
$M_{7}=M/10^{7}M_{\odot}$, $m_{*}=m/M_{\odot}$ and $R_{*}=R_{st}/R_{\odot}$. Since the semi-major
axis of the stellar orbit can be as large as
a few $pc$, tidally disrupted stars have highly elongated (technically, parabolic) orbits. 
Another important process that competes with
the process of the tidal disruption is the relativistic 
process of direct capture of stars by a black
hole. The star can be directly captured by the black hole when its periastron is of the order
of the black hole gravitational radius
\be r_{g}={2GM\over c^{2}}\approx 3\cdot 10^{12}M_{7}cm. \label{eq i2}\ee  
Since for a black hole of mass larger than $10^{7}M_{\odot}$ the ratio $r_{g}/r_{T} > 0.2$,  
the relativistic treatment of the process of the tidal disruption is important.

We consider  problem of  tidal
disruption of a star, or a partial stripping of mass from the star,
by the black hole. Our treatment of the problem is fully based on General Relativity.
We determine numerically an amount of mass lost by the star as a result of
a tidal encounter, for different parameters of the stellar 
orbit and of the black hole. Since the initial orbit is assumed to be parabolic and 
the initial state of the star is assumed to be unperturbed by tidal influence of the black hole,
the terminology associated with cross sections is used, see also Beloborodov et al.
 (1992) (BIIP) and below. As we have mentioned above (and 
will show quantitatively later on), the relativistic effects are not
important when the black hole mass is sufficiently small. On the other
hand when the black hole mass is sufficiently large
$\sim 10^{8}M_{\odot}-10^{9}M_{\odot}$ the stars can be directly
swallowed by the black hole without tidal disruption (see Section
3.1.1 and also BIIP and references therein).
We obtain numerically approximate expressions
for the amount of mass lost by the star for an intermediate mass range of the black hole,  
where the effects of General Relativity are important and the
processes of tidal disruption and tidal stripping of mass 
compete with the process of direct capture.

The rotating black hole can be characterised by only two parameters, its mass $M$ and the 
rotational parameter $a=cL_{h}/(GM^{2}) < 1$, where $L_{h}$ is the black hole angular momentum.

To describe the orbit of the star we must specify the orbit's integrals of motion.
We introduce the spherical coordinate system $(r, \theta, \phi)$ with $\theta=\pi/2$
at the black hole equatorial plane and consider the motion of the star very far from the 
black hole, $r \rightarrow \infty$. In this case  
a highly elongated  orbit of the star can be characterised by the
polar angle $\theta_{\infty}$
associated with the semi-major axis of the orbit, 
and by two projections of specific angular
momentum, $j_{\phi}$ and $j_{\theta}$
\footnote{Let the velocity components at the orbit apastron be
$v_{\phi}$ and $v_{\theta}$ and the apastron distance be $r_{a}$. Then, $j_{\phi}=r_{a}v_{\theta}$
and $j_{\theta}=r_{a}v_{\phi}$.}.  
By definition, these quantities are conserved during the
motion of the star.
Taking into account that the problem is symmetric with
respect to reflection $j_{\phi} \rightarrow -j_{\phi}$ we can consider only positive values of
$j_{\phi} > 0$. 
Alternatively, one can characterise the stellar orbit by projection 
of the specific angular momentum onto the axis of rotation of the black
hole, $L_{z}$ and
the square of the projection of the angular momentum onto the equatorial
plane, $Q$. It is easy to show that the so-defined integral of motion $Q$ 
coincides  with the well-known Carter integral when the
gravitational field of the black hole dominates over the gravitational
field of the central stellar cluster.  
The quantities $L_{z}$ and $Q$ are related to $j_{\phi}$ and $j_{\theta}$ as
\be L_{z}=j_{\theta}\sin \theta_{\infty}, \quad Q=j_{\phi}^{2}+j_{\theta}^{2}\cos^{2} \theta_{\infty}.
\label{eq i3}\ee
The Carter integral $Q$ is always positive for  parabolic orbits, and we will use its square root
$q=\sqrt {Q}$  to have the same dimensions for the integrals of motion $q$ and $L_{z}$, 
$j_{\phi}$, $j_{\theta}$. For $\theta_{\infty}=\pi/2$ and $j_{\phi} > 0$, 
$L_{z}=j_{\theta}$ and $q=j_{\phi}$.   

It is well known that for a given $\theta_{\infty}$ there is a region, $S_{c}$ 
in the plane $(j_{\theta}, j_{\phi})$
corresponding to capture of the stars by the black hole (e.g. Chandrasekhar
1983, Young 1976). By analogy,
we define the cross section of the tidal disruption, $S_{T}$ 
as a region in the plane $(j_{\theta}, j_{\phi})$ where a partial tidal stripping of mass from the 
star or a full tidal disruption of the star takes place (BIIP). An alternative definition of the
tidal cross sections would be a region in the plane $(L_{z}, q)$ for a given $\theta_{\infty}$ 
where a tidal stripping or a tidal disruption takes place. We will call these
as the cross sections in the plane  $(L_{z}, q)$. When $\theta_{\infty}=\pi/2$ the cross sections
in the upper half plane  $(j_{\theta}, j_{\phi} > 0)$ 
and the cross sections in the plane $(L_{z}, q)$ coincide.
It is very important to note that the cross
section of direct capture of the stars in the plane $(L_{z}, q)$ does not depend on the angle
 $\theta_{\infty}$ (e.g. Chandrasekhar 1983)
\footnote{Technically, it follows from the fact that the radius of periastron of a parabolic orbit
 does not depend on  $\theta_{\infty}$.}. As we will see (Section 3.1.3) this property is also
approximately valid for the cross sections of the tidal disruption in the plane $(L_{z}, q)$ (see 
the text below). This fact allows us to reduce the dimensionality of the parameter space and obtain
the approximate cross sections of tidal disruption for $\theta_{\infty} \ne \pi/2$ by a 
geometric transform of the cross sections calculated for  $\theta_{\infty}=\pi/2$.      

In order to calculate the cross sections of tidal disruption one should evolve numerically
a model of a tidally disrupted star in the relativistic tidal field of the
black hole for a sufficiently long time. It should be done
for different values  $j_{\phi}$, $j_{\theta}$, $\theta_{\infty}$ as well as for 
different values of $M$ and $a$. Approximately $10^{3}-10^{5}$ runs are needed.
Taking into account that the present day numerical 3D finite difference
or SPH models are rather time consuming it is impossible to survey the whole parameter space with 
these models
\footnote{See Ivanov $\&$ Novikov (2001) for an overview of works on tidal disruption and
astrophysical applications.}. Recently 
we have proposed another approximate model of a tidally disrupted star
which is a one dimensional Lagrangian model from the point of view of numerical calculations, and
therefore, it is numerically fast (Ivanov $\&$ Novikov 2001, hereafter IN, Ivanov, Chernyakova 
$\&$ Novikov 2003, hereafter ICN). This model is used in the present Paper.
 
The model is briefly described in the next Section.
 Section 3 presents  the results of our calculations.
 We discuss our results and present our conclusions in Section 4.   
 
We use natural units expressing all quantities related to the star in terms of the characteristic 
stellar time $t_{st}=\sqrt {R_{st}^{3}/Gm}$, stellar radius $R_{st}$, etc. The parameters related to
the stellar orbit are also made dimensionless. We use $\tilde L_{z} =L_{z}/(cr_{g})$, 
$\tilde q =q/(cr_{g})$, $\tilde j_{\phi} =j_{\phi}/(cr_{g}) $ and $\tilde j_{\theta}=j_{\theta}/(cr_{g})$.
We do not write tilde below assuming that all quantities mentioned above are dimensionless.

We assume below that the star is a full polytrope with $n=1.5$ and  solar mass
and radius. We show how to apply our
results to the case of other values of the mass and radius.

We mainly consider the tidal encounters of  moderate
strength, thus neglecting the  possibility of violent tidal disruption and
formation of strong shocks in 
the stellar gas.  

\section{The model of a tidally disrupted star}

The derivation of the basic dynamical equations of our model 
 and detailed comparison of the results with
results based on the 3D finite difference models, SPH model and affine models of the
tidally disrupted star has been intensively discussed in our previous papers 
(see IN, ICN). 

We assume 
that summation is performed over
all indices appearing in our expression more than once, but summation is not performed
if indices are enclosed in brackets. The upper and lower indices describe the rows and
columns of matrices.

In our model, in order to reduce the number of degrees of freedom, and accordingly, to
 significantly reduce the computational time we assume that the star consists of 
elliptical shells. The parameters and orientation of these shells change with time. 
This assumption 
allows us to express the coordinates of some particular gas element in a frame co-moving 
with the centre of mass of the star, $x^{i}$ in terms of (non-evolving) coordinates of the
same gas element $x^{i}_{0}$ in the unperturbed spherical state of the star (say,
before the tidal field is 'switched on') as    
\be
x^{i}=T^{i}_{j}(t, r_{0})e_{0}^{j},\label{eq 1}
\ee 
where the radius $r_{0}=\sqrt{x_{0i}x^{i}_{0}}$ is the same for all gas element 
belonging to a particular
shell, and 
$e_{0}^{i}=x^{i}_{0}/r_{0}$ are direction cosines 
($e_{0i}e^{i}_{0}=1$).
We represent
the position matrix $T^{i}_{j}$ and its inverse
$S^{i}_{j}$ as a product of
two rotational matrices $A^{i}_{j}$ and $E^{i}_{j}$, and a diagonal 
matrix $B^{i}_{j}$:
\be T^{i}_{j}=A^{i}_{l}B^{l}_{m}E^{m}_{j}=a_{l}A^{i}_{l}E^{l}_{j}, 
\quad S^{i}_{j}=a_{l}^{-1}A^{j}_{l}E^{l}_{i}, \label{eq 2} \ee
where $B^{l}_{m}=a_{(l)}\delta^{(l)}_{m}$, and $a_{l}$ are the principal
axes of the elliptical shell.  

The evolution equations for the matrix $T^{i}_{j}$ follow from the integral 
consequences of the exact hydrodynamical of motion: the so-called 
virial relations (e.g. Chandrasekhar, 1969) 
written for a particular elliptical shell of interest, and have the form:
\[ \ddot T^{i}_{j}=3S^{j}_{i}\bar {({p\over \rho})}-12\pi S^{j}_{k}{d\over dM}
\lbrace gS^{l}_{i}T^{k}_{n}{\bar P}^{ln}\rbrace \]
\be -{3\over 2}A^{i}_{k}a_{k}D_{k}E^{k}_{j}{GM\over g}+C^{i}_{k}T^{k}_{j},
\label{eq 3} \ee 
where the dot stands for the time derivative. 
We use the mass enclosed in the shell corresponding to
the radius $r_{0}$, $M=4\pi \int_{0}^{r_{0}}\rho_{0}r_{1}^{2}dr_{1}$ as a new
radial variable, and $\rho_{0}$ is the density of the star in its unperturbed
spherical state. The matrix $S^{i}_{j}$ represents the inverse to the 
position matrix $T^{i}_{j}$ and $g$ is the determinant of the position matrix.
The matrix $C_{i}^{j}$ represents the relativistic tidal tensor, and therefore
it is symmetric and traceless.

The dimensionless
quantities $D_{k}$ have been described by e. g. Chandrasekhar (1969), p. 41.
They have the form:
\be D_{k}=g\int^{\infty}_{0}{du\over \Delta (a_{j}^{2}+u)}, \label{eq 3a} \ee
where $\Delta=\sqrt{(a_{1}^{2}+u)(a_{2}^{2}+u)(a_{3}^{2}+u)}$.
 
The quantities $\bar {({p\over \rho})}$ and ${\bar P}^{ij}$ are determined by
distribution of the density $\rho $ and pressure $p$ in the star:
\be \bar {({p\over \rho})}={1\over 4\pi}\int d\Omega {p\over \rho}, \label{eq 4}\ee
and 
\be {\bar P}^{ij}={1\over 4\pi}\int d\Omega pe_{0}^{i}e_{0}^{j}, \label{eq 5} \ee
where integration is performed over the unit sphere associated with the direction
cosines $e_{0}^{i}$.

Obviously, the first two terms in equation \ref{eq 3} are due to action of the pressure
force on the elliptical shell, and the last two terms describe the self-gravity
of the star and action of the tidal forces.  

To complete the set of equations \ref{eq 3} we should know the distributions of pressure
and density over the star. The density  distribution 
\be \rho(t,x^{i})={\rho_{0}(r_{0})\over D} \label{eq 6} \ee
is determined by the Jacobian 
$D=|{\partial x^{i} \over \partial x_{0}^{j}}|$ of the
mapping between the coordinates $x^{i}$ and $x_{0}^{i}$, which can be written as
\be D(x_{0}^{i})={ge_{0i}e_{0k}\over
2 r_{0}^{2}} (S^{i}_{m}(T^{m}_{k})\prime + S^{k}_{m}(T^{m}_{i})\prime, \label{eq 7} \ee
where prime stands for differentiation over $r_{0}$. We use the standard relation
$p=k\rho^{\gamma}$ to find the distribution of pressure over the star. Here $k$ is the entropy
constant, and $\gamma$ is the specific heat ratio. $\gamma=5/3$ is used later on.

The dynamical equation  \ref{eq 3} has the usual integrals of motion: the energy integral 
and the integral of angular momentum. Their form can be found in ICN. For
our discussion it is important that in our model the energy integral 
can be naturally separated into  kinetic,
thermal and gravitational parts. In addition to these integrals, the quantities
\be \chi_{jk}(M)=T^{i}_{k}\dot T^{i}_{j}-T^{i}_{j}\dot T^{i}_{k} \label{ eq 9} \ee
are conserved. They represent the conservation of circulation of the fluid over the elliptical 
shells in our model (IN, ICN). 

\subsection{Numerical scheme and computational details}

Our numerical scheme is described in IN and ICN. The variant of the scheme used in
the calculations is a non-conservative explicit numerical scheme, and we use conservation of the
integrals of motion to check the accuracy of the calculations. This simple scheme allows us
to calculate a single tidal disruption event with a very small computational time. However, as 
was pointed out in ICN, this variant of the scheme suffers from a slowly growing numerical
instability. A radical remedy would be an implicit conservative numerical
scheme, but the schemes of this type are much more time
consuming. Therefore, in order to suppress this instability we 
average the dynamical variables over the neighboring grid points 
once every $200$ time steps. We have checked that this 
procedure leads only to a rather slow numerical leakage of the stellar energy  
which seems to be non-significant for our purposes.

We use a simple criterion of  mass loss. We calculate the sum of kinetic and potential energy
(per unit of mass) for each grid point and assume that  when this sum
is positive the corresponding elliptical shell is 
gravitationally unbound. This criterion is in  good agreement with
the results of 3D finite difference computations (see IN and ICN).

In principal one can use other criteria of the amount of mass
lost by the star. For example, one can consider the shells with
positive total energy (i. e. the sum of the potential, kinetic and
thermal energies) as being tidally stripped from the star. For a
particular shell the ratio
of the thermal energy $E_{th}$ to the potential energy $E_{p}$ is
proportional to a power of a characteristic size of the shell,
$R_{sh}$: $E_{th}/E_{p} \propto R_{sh}^{-(3\gamma -4)}$. 
When $\gamma > 4/3$ this ratio decreases with time for an expanding
shell. Therefore, taking into account the thermal energy should not make
a significant difference to our criterion. 

 Our model cannot treat 
shocks which can, in principle, influence the criterion of the mass
loss. However,  we mainly consider tidal encounters
of  moderate strength where only a partial mass stripping of the star
takes place and the presence of strong shocks is not expected. This
statement is confirmed by results of 3D finite difference
calculations where no evidence for a strong shock in the star has been 
found (e.g. Khokhlov et al. (1993), Diener et al. (1997)). Note that
some shocks may  develop after the tidal disruption event in a
situation when  partial mass stripping takes places. A gravitationally bound
part of the stellar gas initially expelled from the star may return and
hit the surface of a dense undisrupted stellar core producing a shock
there. These shocks only redistribute the energy and momentum of the
gravitationally bound part of the stellar gas and therefore cannot
influence our criterion.

For the calculation of the stellar orbits we use 
the usual Boyer-Lindquist coordinates $(r, \theta, \phi)$
and the explicit form of the tidal tensor given in Diener et al. (1997)
\footnote{Note  misprints in the Diener et al. (1997) 
expression for the components of the tidal tensor.
The sign of all components should be opposite.}.  

As follows from the results based on the finite difference 
scheme (e.g Diener et al. 1997, and references
therein) and our own results (IN, ICN) the stellar 
structure starts to evolve significantly only
when the star approaches the periastron of the orbit, $r_{p}$. 
Therefore we start our calculations at the radius
$r_{0}=1.5r_{g}$ and evolve our numerical scheme during the fly-by
around the black hole
until a final sufficiently large
radius $r_{fin}$ is reached. This radius is taken to be $r_{fin}=4-5r_{p}$ 
for a sufficiently small mass of the black hole, $M < 8\cdot 10^{7}M_{\odot}$. In the opposite case
we terminate a single computation when two conditions are satisfied: 1) $r_{fin} > 3r_{g}$; 
2) the dimensionless time $\tau=t/t_{*}$ from the 
beginning of computation is sufficiently large:
$\tau > 15$.     
   
\begin{figure}
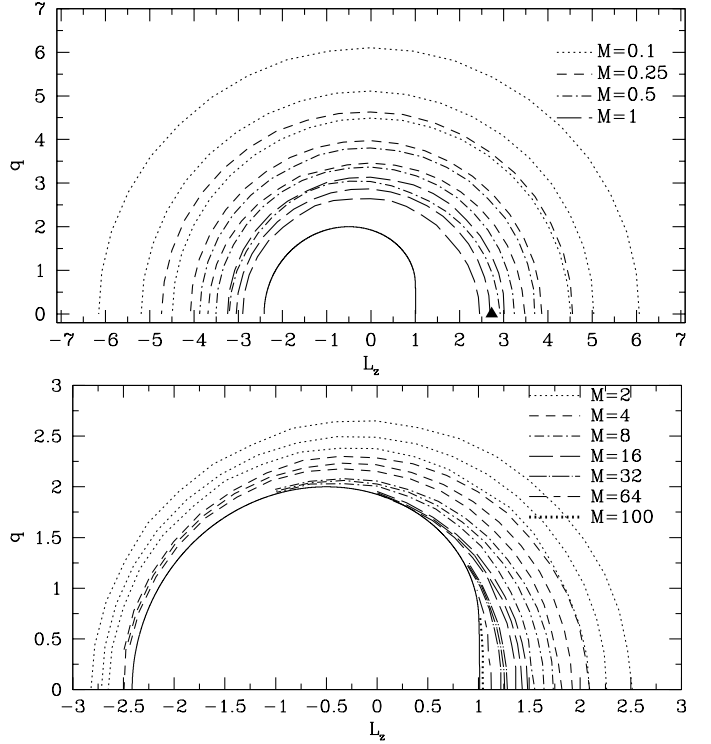

\begin{center}
\includegraphics[width=9cm,angle=0]{f1.epsi}
\includegraphics[width=9cm,angle=0]{f2.epsi}
\end{center}
\caption{The calculated cross sections of tidal disruption for the case 
$a=0.999$ and different black hole masses. 
The levels of the amount of mass lost by the star, $M_{lost}$, are shown by curves 
of different types corresponding to different masses of the black hole. 
Three curves of the same type show $M_{lost}=0.1$ (the outer curve),
$M_{lost}=0.5$ (the middle curve) and $M_{lost}=1$ (the inner curve).
Note that the curves corresponding to the different black hole masses 
may overlap (e.g. $M_{lost}(M=0.1)=1$ and $M_{lost}(M=0.25)=0.1$). 
The solid curve
shows the cross section of direct capture.
{\bf \it Top}: The case of small black hole masses is shown, $M=0.1$, $0.25$, $0.5$ and $1$.
The triangle shows the result of 3D finite difference calculations of Diener et al. (1997).
They obtained $M^{D}_{lost} \sim 0.5$.
{\bf \it Bottom}: The case of large black hole masses, $M=2$, $4$, $8$, $16$, $32$, $64$ and 
$100$. Note that in the case $M=64$ and $M=100$ 
the curves corresponding to different $M_{lost}$ almost
coincide,  and we show only the curves corresponding to $M_{lost}=1$.}
\label{theor}
\end{figure}
\begin{figure}
\begin{center}
\includegraphics[width=9cm,angle=0]{f3.epsi}
\includegraphics[width=9cm,angle=0]{f4.epsi}
\end{center}
\caption{Same as Fig. 1 but $a=0.75$ {\bf \it Top}: $M=0.1$, $0.25$, $0.5$ and $1$
 {\bf \it Bottom}: $M=2$, $4$, $8$, $16$, $32$}
\label{theor1}
\end{figure}\begin{figure}
\begin{center}
\includegraphics[width=9cm,angle=0]{f5.epsi}
\includegraphics[width=9cm,angle=0]{f6.epsi}
\end{center}
\caption{Same as Fig. 1 but $a=0.5$ {\bf \it Top}: $M=0.1$, $0.25$, $0.5$ and $1$
 {\bf \it Bottom}: $M=2$, $4$, $8$, $16$}
\label{theor2}
\end{figure}\begin{figure}
\begin{center}
\includegraphics[width=9cm,angle=0]{f7.epsi}
\includegraphics[width=9cm,angle=0]{f8.epsi}
\end{center}
\caption{Same as Fig. 1 but $a=0.25$ {\bf \it Top}: $M=0.1$, $0.25$, $0.5$ and $1$
 {\bf \it Bottom}: $M=2$, $4$, $8$}
\label{theor3}
\end{figure}
\begin{figure}
\begin{center}
\includegraphics[width=9cm,angle=0]{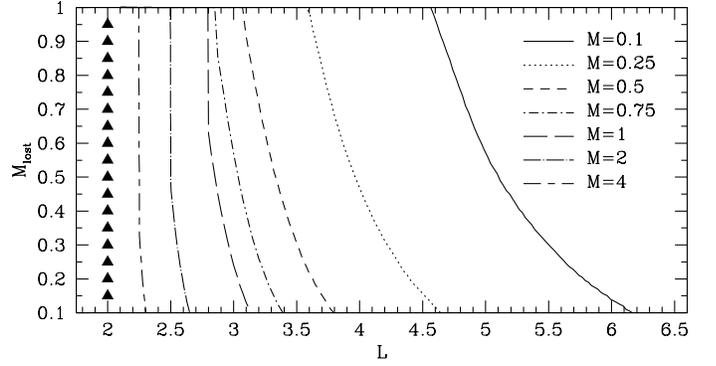}
\end{center}
\caption{The amount of mass lost by the star calculated 
for the case $a=0$. Since the cross sections in this case are 
concentric circles in the plane $(L_{z}, q)$ we show $M_{lost}$ as a function of the radius
$L=\sqrt{q^{2}+L_{z}^{2}}$ for the different black hole masses $M=0.1$, 
$0.25$, $0.5$, $0.75$, $1$, $2$ and
$4$. Vertical triangles show the position 
of the radius of the capture cross section $L_{capt}=2$.}
\label{ml0}
\end{figure}

\begin{figure}
\begin{center}
\includegraphics[width=9cm,angle=0]{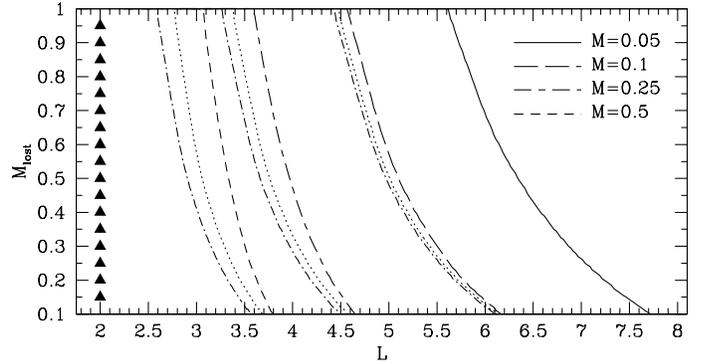}
\end{center}
\caption{The amount of mass lost by the star  compared with the 
results based on the Newtonian approach to the problem for the case $a=0$.The dash 
dotted curves correspond to extrapolation
to the higher masses of the results obtained for $M=0.05$ and the dotted curves take into account
the relativistic dependence of the periastron radius $r_{p}$ on $L$, see the text for details.}
\label{ml0comp}
\end{figure}

\section{Results} 

\subsection{The case $\theta_{\infty}=\pi/2$}

The bulk of our computations has been performed for $\theta_{\infty}=\pi/2$. In this case 
the tidal cross sections in the upper half plane  $(j_{\theta}, j_{\phi} > 0)$ and
the cross sections in the plane $(L_{z}, q)$ coincide. For symmetry reasons we can consider 
only  the cross sections in the plane $(L_{z}, q)$.
As it is shown later
(see Section 3.3) the tidal cross sections in the plane $(L_{z}, q)$ almost do not depend
on the value of $\theta_{\infty}$ and therefore the cross sections in the plane
$(j_{\phi}, j_{\theta})$ can be obtained from those corresponding to $\theta_{\infty}=\pi/2$ by a 
geometric transform.  

\subsubsection{The form of cross sections}

In order to calculate the tidal cross sections 
we  specify the range of the black hole masses and rotational parameters. We 
calculate the amount of mass lost by the star, $M_{lost}$, for 
a given value of the black hole mass and rotational parameter. These values are
shown in Table 1. We plot the levels of constant $M_{lost}$ in the plane  $(L_{z}, q)$.
The black hole mass $M$ is expressed in
units of $10^{7}M_{\odot}$ and  $M_{lost}$ is expressed 
in units of the stellar mass $m_{*}$ in all Figures. 
\begin{table*}
 \centering
 \begin{minipage}{140mm}
  \caption{The values of the black hole masses and rotational parameters used in the 
   calculations}
  \begin{tabular}{@{}llrrrrlrlr@{}}
      
 \hline
   $M/ 10^{7} M_{\odot}$ & 0.05 & 0.1 & 0.25 & 0.5 & 1 & 2\\  
   $M/ 10^{7} M_{\odot}$ & 4 & 8 & 16 & 32 & 64 & 100 & \\
   $a$ & 0 & 0.25 & 05 & 0.75 & 0.9 & 0.999 \\

\hline
\end{tabular}
\end{minipage}
\end{table*}

In Figures 1-4 we show the cross sections for $a\ne 0$. Three levels of  $M_{lost}$=$0.1$,
$0.5$ and $1$ are shown for a given $a$ and $M$. Since  
the cross section of direct
capture does not depend on $M$ in our dimensionless units, we plot several different tidal
cross sections and the cross section of direct capture 
corresponding to several different black hole masses in the
same Figure. 

In the non-relativistic theory of tidal disruption the tidal cross sections are circles centred at 
the origin of the coordinates  $(L_{z}, q)$  with radius
$\propto M^{-1/3}$ in our units
\footnote{In  physical units the characteristic size of the tidal
cross section $\propto M^{2/3}$ and the characteristic size of the
capture cross section $\propto M$.}. 
When the mass of the black hole exceeds a certain value
the radius of the tidal cross section becomes smaller than the one corresponding to capture and
the tidal disruption can no longer take place. In the fully relativistic treatment of the
problem the situation is more complicated. We have checked with an accuracy
of the order of $10^{-3}$ that even in this case 
the levels of the mass loss always have a circular form. 
However, when $a > 0$  
the centres of these circles are shifted toward negative values
of $L_{z}$ and the radii of these circles do not obey the simple $\propto M^{-1/3}$ law (see  
Figures 7--11 and the Section 3.1.2).

These effects are most prominent at a high rotational rate of the
black hole. In Figure 1 we show the results of
calculations for $a=0.9999$. When the black hole mass is sufficiently
small (see the upper part of this
Figure), the levels are approximately circular with a small shift
toward negative values of $L_{z}$. 
The triangle shows the position of the orbital parameters 
$L_{z}=2.72945$ and $q=0$ for the case intensively
studied by Diener et al. (1997).  They  obtained $M_{lost}\approx 0.5$ for 
$M=1.0853\times 10^{7}M_{\odot}$. This value is in
 excellent agreement with our results for the mass $M=10^{7}M_{\odot}$. Note, however, that
there is a considerable ambiguity in the results reported by Diener et al. (1997)  counting
the gas elements leaving the computational box but having  velocities less than the parabolic velocity
as being stripped away from the star. Provided that these elements are not considered as being stripped,
the value of $M_{lost}$ is considerably less, $M_{lost}\approx 0.32$. The lower part of Figure 
1 shows the case of high black hole masses. The levels of the amount of mass lost by the star 
decrease toward the 
capture cross section with increasing mass. When 
$M \equiv M_{-} \sim 4\cdot 10^{7}M_{\odot}$ the levels of mass loss intersect the capture cross section 
at the point $(q=0, L_{z}=L_{-})$, where 
\be L_{-}(a)=-(1+\sqrt{1-a}), \label{req 1}\ee 
is a 'critical' negative angular momentum. It is defined for
the equatorial orbits
with negative angular momenta if the particles on the orbits with $|L| <
|L_{-}|$ are captured by the black hole.
When the black hole mass grows further, the intersection point moves toward the positive
value of $L_{z}$ and when $M > 1.6\cdot 10^{8}M_{\odot}$ only the stars with positive angular momenta
can be tidally disrupted. When $M\equiv M_{+}\sim 10^{9}M_{\odot}$ the tidal cross section
coincides with the capture cross section. In this case only the stars with angular momenta
very close to 
\be L_{+}=1+\sqrt{1-a} \label{req 2}\ee
can be tidally disrupted. Here the 'critical' angular momentum 
$L_{+}$ is 
defined analogously to $L_{-}$ for the orbits with positive angular
momenta (i. e. the stars with $L < L_{+}$ are captured by the
black hole). 
When  $M> M_{+}$ only the process of capture of the
stars is possible.  The  width of the region in the plane $(L_{z}, q)$ 
where a partial stripping of mass takes place  decreases with mass. As we will see below this is
mainly a relativistic effect. 

In Figures $2-4$ we show  similar results for smaller values of the rotational parameter,
$a=0.75$ (Figure 2), $0.5$ (Figure 3), $0.25$ (Figure 4).
In general, the behaviour of the levels of $M_{lost}$ 
looks similar to the previous case and the levels of constant $M_{lost}$ can be well 
approximated as circles. The shift of the centres of 
these circles from the origin of the coordinates  
decreases with the decrease of $a$. The corresponding values of $M_{-}$ and $M_{+}$ are 
shown in Table 2. As seen from this Table
the value of $M_{-}$ slightly increases and the
value of $M_{+}$ decreases with decrease of $a$.
\begin{table*}
 \centering
 \begin{minipage}{140mm}
  \caption{The values of $M_{-}$ and $M_{+}$ as functions of $a$ }
  \begin{tabular}{@{}llrrrrlrlr@{}}

 \hline $a$ & 0.25 & 0.5 & 0.75 & 0.999 \\

 \hline
   $M_{-}$ & $\sim 8\cdot 10^{7}M_{\odot}$ & $\sim 8\cdot 10^{7}M_{\odot}$  
    & $< 8\cdot 10^{7}M_{\odot}$ & $\sim 4\cdot 10^{7}M_{\odot}$ \\  
   $M_{+}$ & $\sim 8\cdot 10^{7}M_{\odot}$ & $\sim 1.6\cdot 10^{8}M_{\odot}$  
   & $\sim 3.2\cdot 10^{8}M_{\odot}$   & $\sim 10^{9}M_{\odot}$  \\
   
\hline
\end{tabular}
\end{minipage}
\end{table*}

When $a=0$ the cross sections are circles centred on the origin of the coordinate system,
and we can use the dependence of $M_{lost}$ on $L=\sqrt{L_{z}^{2}+q^{2}}$ to represent our 
results. This is shown in Figure 5. In this case $M_{-}=M_{+}\sim 4 \cdot 10^{7}M_{\odot}$. 
Similary to the case
of non-zero $a$ the size of the region where a partial stripping takes place decreases with 
$M$. To show how important the relativistic effects are
we compare our results with simple
relations for the amount of mass lost by the star 
based on the non-relativistic theory of tidal disruption, see Figure 6. 
In the non-relativistic theory 
$M_{lost}$ depends on the periastron distance $r_{p}$, the black hole mass $M$,
the stellar mass $m_{*}$ and the stellar
radius $R_{st}$ only in combination
\be \eta={(r_{p}/r_{T})}^{3/2}, \label{req 3}\ee
where $r_{T}$ is the tidal radius given by equation \ref{eq i1}. 
We calculate the
dependence $M_{lost}(\eta)$ for a very low mass $M=5\cdot 10^{5}M_{\odot}$,
and use this dependence to calculate $M_{lost}(L)$ for higher masses of the black hole (dot dashed curves
in Figure 6). 
These curves give a good approximation to the relativistic calculations for a 
small black hole mass $M=10^{6}M_{\odot}$. However, the results  diverge with increasing  
mass $M$. The curves based on the non-relativistic approximation systematically give a larger size
of the region where a partial stripping of the mass takes place. We have checked that when 
$M > 10^{7}M_{\odot}$ the deviation is very significant and these results are not shown in
this Figure. 
To take into account some effects of General Relativity, Diener et al. (1997) have proposed using
the relativistic dependence of $r_{p}$ on $L$ in equation \ref{req 3}. The curves calculated according
to this prescription are shown as dotted curves in Figure 6.
The prescription of Diener et al. (1997) slightly improves the agreement, but the deviation
is still rather large. This can be explained by the fact that
there are relativistic effects not accounted for by the Diener et al. (1997) prescription, such as
a different form of the tidal tensor, a different time spent by the star near the periastron, etc..

\begin{figure}
\begin{center}
\includegraphics[width=9cm,angle=0]{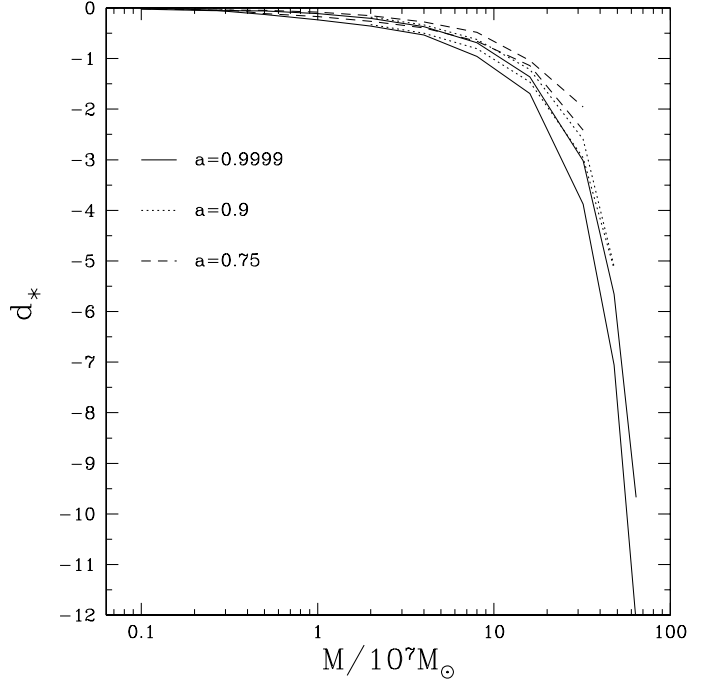}
\end{center}
\caption{The dependence of the shift of the approximating circles $d_{*}$ on mass $M$ for three values of the
rotational parameter $a=0.9999$, $0.9$ and $0.75$. The lower and upper curves of the same type correspond
to the circles approximating the levels  $M_{lost}=1$ and $M_{lost}=0.1$, respectively.}
\label{dmabig}
\end{figure}\begin{figure}
\begin{center}
\includegraphics[width=9cm,angle=0]{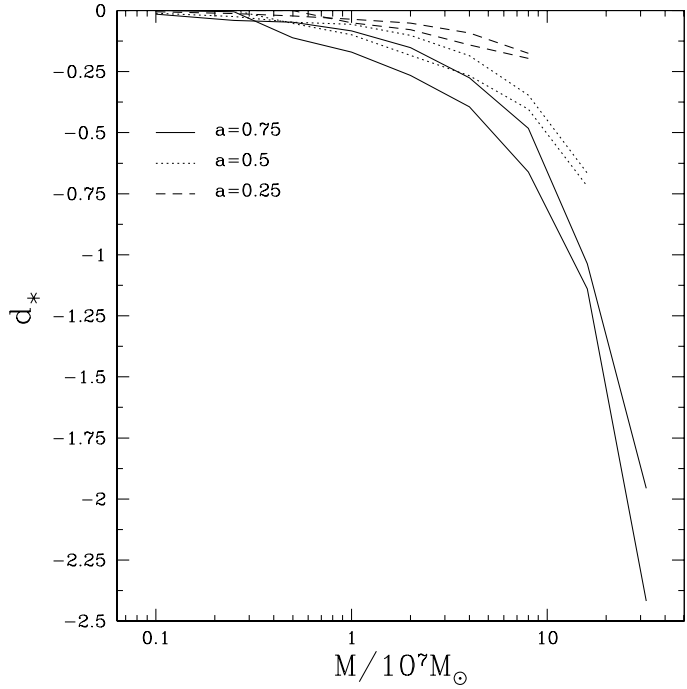}
\end{center}
\caption{Same as Figure \ref{dmabig} but  $a=0.75$, $0.5$ and $0.25$.}
\label{dmasmall}
\end{figure}
\begin{figure}
\begin{center}
\includegraphics[width=9cm,angle=0]{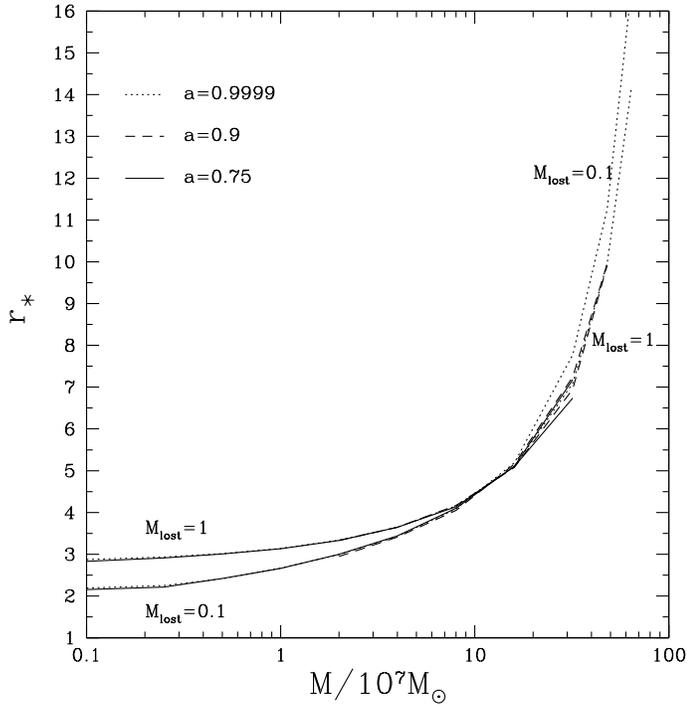}
\end{center}
\caption{The dependence of the radius of the approximating circles $r_{*}$ on mass $M$ for three values of the
rotational parameter $a=0.9999$, $0.9$ and $0.75$. The curves corresponding to the same value
of $a$ and describing two different values of the amount of mass lost by the star 
$M_{lost}=0.1$ and $M_{lost}=1$ intersect 
at $M\sim 10$. For black hole higher masses the radius of the circle corresponding to 
$M_{lost}=1$ has a radius $r_{*}$ larger than for $M_{lost}=0.1$.}
\label{rm}
\end{figure}
\begin{figure}
\begin{center}
\includegraphics[width=9cm,angle=0]{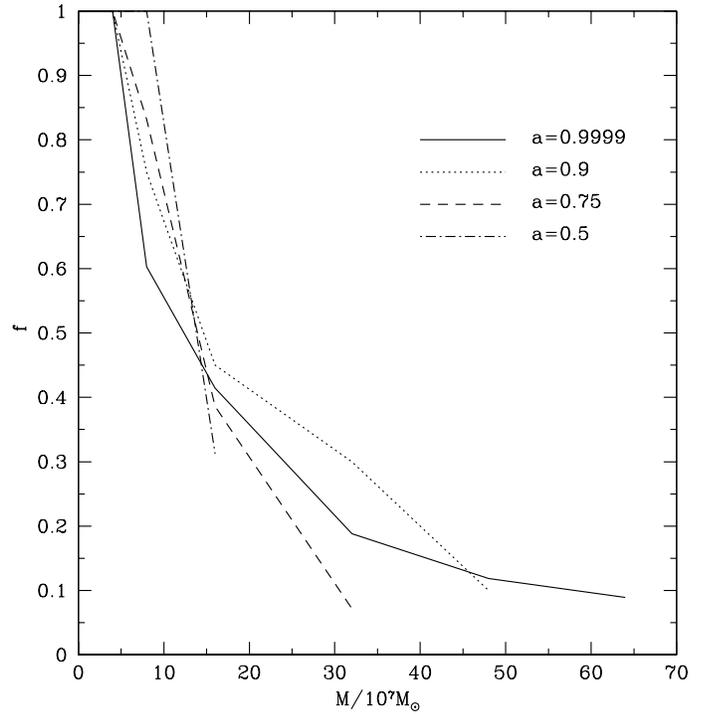}
\end{center}
\caption{The ratio $f$ of the length of a segment of the approximating
circle lying outside the capture cross section to
the total length of the approximating circle as a function
of the mass $M$. All curves correspond to $M_{lost}=0.1$ 
and different values of $a=0.9999$, $0.9$, $0.75$ and
$0.5$.}
\label{fm}
\end{figure}
\begin{figure}
\begin{center}
\includegraphics[width=8cm,angle=0]{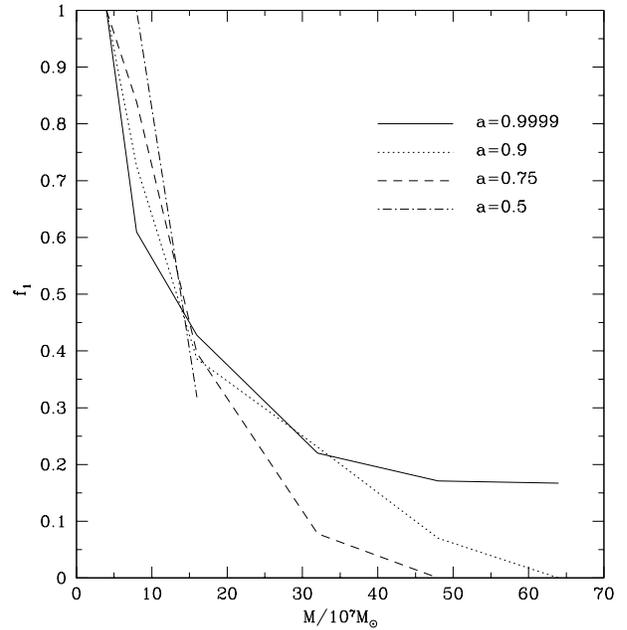}
\end{center}
\caption{Similar to Figure 10 but now the ratio $f_{1}$ of the length of an approximating 
circle lying outside the capture cross section to the total length of the 
unit of the tidal cross section and
the capture cross section is shown as a function of the mass $M$.}
\label{f1m}
\end{figure}

\subsubsection{Semi-analytic representation of the results obtained}

As we discussed above the contours of the amount of mass lost by the star 
in the plane $(L_{z}, q)$ 
can be approximated as circles of a given
radius $r_{L}$ shifted toward negative values of $L_{z}$. This means that they can
be represented in the form:
\be (L_{z}-d)^{2}+q^{2}=r_{L}^{2}, \label{req 4}  \ee
where $d < 0$ is the shift of the centres of the approximating circles. 
Therefore,
we can describe the levels of $M_{lost}$  
with  two functions $d(M,a)$ and $r(M,a)$.
We present these functions
in Figures 7-11 for two values of $M_{lost}$: $M_{lost}=0.1$ and $M_{lost}=1$. 
Since in the non-relativistic approximation 
the characteristic size of the cross sections scales as
$M^{-1/3}$, it is convenient to introduce the rescaled radius and shift:
$d_{*}=M^{1/3}d$ and $r_{*}=M^{1/3}r_{L}$. The change of $r_{*}$ with
mass is a purely relativistic effect.

In Figure 7 we present the dependence of $d_{*}$ with mass for high values of the 
rotational parameter $a=0.9999$, $0.9$ and $0.75$.  For any given
curve the rightmost value of $M$ is equal to $M_{+}$.
For a given value of $a$ the shift corresponding to $M_{lost}=1$ is always larger 
than that corresponding to $M_{lost}=0.1$.
In  Figure 8 we show the same results calculated for 
the case of the small rotational parameters $a=0.75$, $a=0.5$ and $a=0.25$.

In Figure 9 we show the dependence of $r_{*}$ on mass for $a=0.9999$, 
$0.9$ and $0.75$. Note that all curves are very close to each
other. The radius
corresponding to the level $M_{lost}=0.1$ is larger than that corresponding to $M_{lost}=1$
only for sufficiently small masses 
$M < M_{intersect} \sim 10^{8}M_{\odot}$.
When $M=M_{intersect}$
the curves corresponding to the different levels intersect  and when 
$M > M_{intersect}$ the radius of the cross section corresponding to $M_{lost}=1$ is larger
than that corresponding to $M_{lost}=0.1$. However, in the region
outside the capture cross section the level corresponding
to $M_{lost}=1$ is situated inside the level corresponding to $M_{lost}=0.1$
due to a larger negative shift. When $M < M_{intersect}$, the curves with
different $a$  coincide. When $a < 0.5$, we have
$M_{+} < M_{intersect}$ and the curves corresponding to smaller values of $a$ 
coincide with the curves corresponding to the larger values of $a$. Therefore,  
we do not show these curves in Figure 9. 

In Figure 10 we show the ratio $f$ of  the length of the circles lying outside the capture
cross section to the total length of the circles. 
The level $M_{lost}=0.1$ is shown for
the cases $a=0.9999$, $0.9$, $0.75$ and $0.5$. For a particular
curve, the mass of the black hole 
corresponding to $f=1$ is approximately equal to $M_{-}$ and the mass
corresponding to the end of a particular curve is approximately  $M_{+}$.
When $a < 0.5$,  $M_{-}$ almost coincides with $M_{+}$ and the curves are
almost vertical. They are not shown in this Figure.

Figure 11 is similar to Figure 10, but here we show the ratio $f_{1}$ of length 
of the circles lying outside the capture cross section 
to the length of the boundary of unit of 
the tidal cross section and the capture cross section. 
This quantity has a direct physical
meaning. In a real astrophysical setting where the super-massive black hole is embedded in a
stellar cluster, the processes of tidal disruption and capture of the stars typically 
occur in the so-called regime
of the empty loss cone (e.g. Frank $\&$ Rees (1976), BIIP and references
therein). In this regime
only the stars with  orbital parameters close to the boundary of the unit
can be either disrupted or captured. So, the quantity $f_{1}$ gives the fraction of tidally
disrupted stars among all stars with $\Theta_{\infty} \approx \pi/2$
that have been destroyed by the black hole (i.e. either tidally disrupted or captured). 
\begin{figure}
\begin{center}
\includegraphics[width=9cm,angle=0]{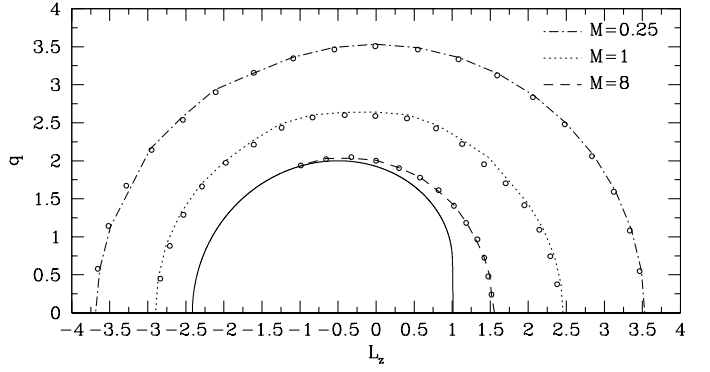}
\end{center}
\caption{The tidal cross sections calculated for $\Theta_{\infty}=\Theta_{+}$ (shown by the circles)
in comparison to
the tidal cross sections calculated for $\Theta_{\infty}=\pi/2$ for $a=0.9999$ and 
$M_{lost}=1$. Note that the case $\Theta_{\infty}=\Theta_{-}$
is equivalent to the case shown, by symmetry.}
\label{tinfp2}
\end{figure}
\begin{figure}
\begin{center}
\includegraphics[width=9cm,angle=0]{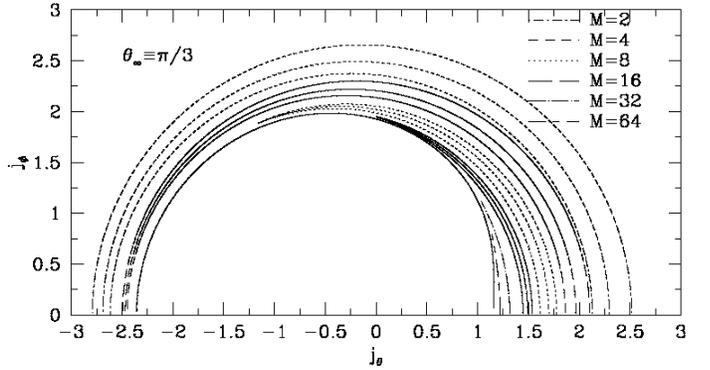}
\end{center}
\caption{The tidal cross sections in the plane $(j_{\theta}, j_{\phi})$. The angle 
$\Theta_{\infty}=\pi/3$ and $a=0.9999$.}
\label{tinfp3}
\end{figure}
\begin{figure}
\begin{center}
\includegraphics[width=9cm,angle=0]{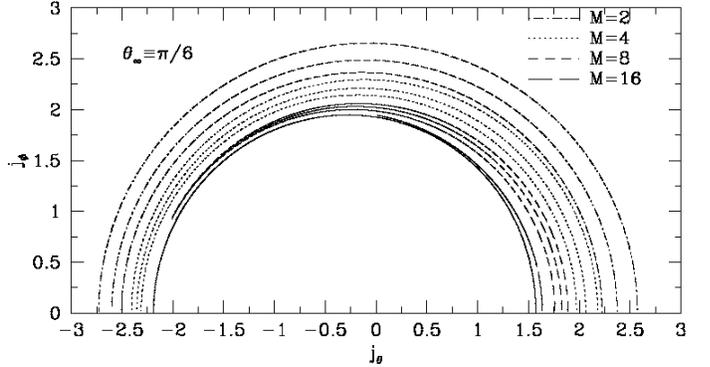}
\end{center}
\caption{Same as Figure 13 but $\Theta_{\infty}=\pi/6$. }
\label{tinfp6}
\end{figure}
\begin{figure}
\begin{center}
\includegraphics[width=9cm,angle=0]{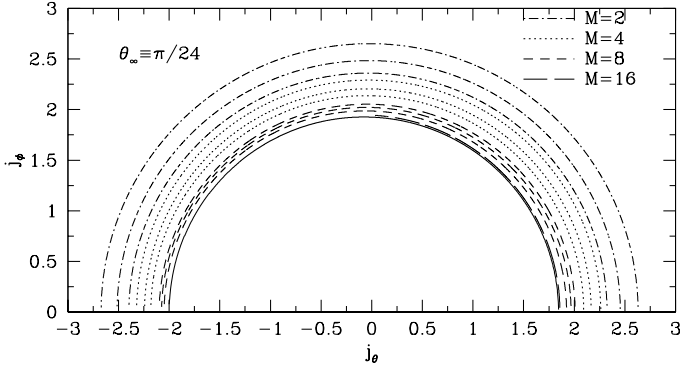}
\end{center}
\caption{Same as Figure 13 but $\Theta_{\infty}=\pi/12$.}
\label{tinfp12}
\end{figure}
\begin{figure}
\begin{center}
\includegraphics[width=9cm,angle=0]{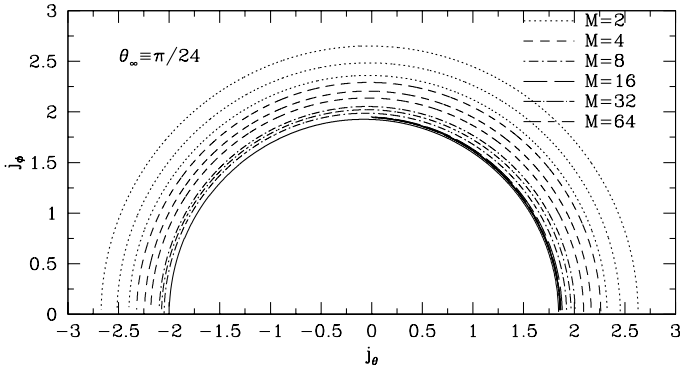}
\end{center}
\caption{Same as Figure 13 but $\Theta_{\infty}=\pi/24$.}
\label{tinfp24}
\end{figure}  

\subsubsection{The case $\Theta_{\infty }\ne \pi/2$}

As follows from equation \ref{eq i3} the law of transformation from the coordinates $(L_{z}, q)$
to the coordinates $(j_{\theta}, j_{\phi})$ has the form
\be  j_{\theta}=L_{z}/\cos \Theta_{\infty}, \quad 
j_{\phi}=\sqrt{q^{2}-L_{z}^{2}\cot^{2} \Theta_{\infty}}. \label{req 5}  \ee                               
Since the  expression in the square root in equation \ref{req 5} must be positive,
only the values of $\Theta_{\infty}$ in the range
\be \Theta_{-}\leq \Theta_{\infty} \leq \Theta_{+} \label{req 6}\ee
are allowed
\footnote{Note that these inequalities are also valid for the angle
$\Theta(\tau)$ that changes during the fly-by around the black hole: 
$\Theta_{-}\leq \Theta (\tau) \leq \Theta_{+}$.}, where
\be \Theta_{\pm}=\pi/2 \pm \arctan {|{q\over L_{z}}|}. \label{req 7} \ee 

In order to check the dependence of our results on $\Theta_{\infty}$ we calculate several tidal
cross sections in the plane $(L_{z}, q)$ for $\Theta_{\infty}=\Theta_{+}$ and compare them to those
corresponding to $\Theta_{\infty}=\pi/2$. The results of comparison are presented in Figure
12. They show that this dependence is practically absent. This may be explained as follows. 
When the periastron distance $r_{p}$ is sufficiently large, the effects
determined by the rotation of the black hole, such as the dependence of the results on 
$\Theta_{\infty}$, are determined by high order corrections 
to the non-relativistic results. In the 
opposite case of a star that has the orbital parameters very close to those corresponding 
to the capture, the orbit of the star near the periastron
is close to a 'barrel-like' critical orbit of
constant $r_{p}$. In this case, the angle $\Theta$  changes between
two limiting values $\Theta_{-}$ and $\Theta_{+}$ near the periastron,
and the dependence of the results on $\Theta_{\infty}$ is averaged out
(see also BIIP).
Therefore, this dependence appears to be unimportant in the
two limiting cases. The 
results shown in Figure 12 indicate that this dependence is also small in an intermediate
case.
 
Therefore, for a given value of $\Theta_{\infty}$, we 
obtain the tidal cross sections in the plane $(j_{\theta}, j_{\phi})$
transforming the results obtained for
$\Theta_{\infty}=\pi/2$ to the new coordinates according
to the transformation law \ref{req 5}. The results are shown in Figures 13--16. 
One can see from these Figures that the cross sections becomes more symmetric with respect
to the origin of the coordinate system with decreasing of $\Theta_{\infty}$. In the limiting
case $\Theta_{\infty}=0$ the cross sections are circles. 

Let us consider the situation when the tidal cross section and the capture cross
section intersect. The relative position of these cross sections is  mainly
characterised by their intersection point. Let $L_{z}^{int}$ be the value of the $z$ component 
of the angular momentum corresponding to the point of intersection on the plane $(L_{z},q)$,
and $j_{\theta}^{int}$ be the coordinate of the same point on the plane $(j_{\theta}, j_{\phi})$.
The values of $L_{z}^{int}$ and $j_{\theta}^{int}$ are connected by equation 17.
As follows from this equation, when $L^{int}_{z} < 0$, $j_{\theta}^{int}$
shifts leftward with decreasing of $\Theta_{\infty}$ from
$\pi/2$ to $0$. In the opposite case $L^{int}_{z} > 0$, $j_{\theta}^{int}$ 
shifts to the right. As follows from Figure 1, we have  
$L_{z}^{int}\approx 0$ for $M\approx 1.6\cdot 10^{8}M_{\odot}$, and as
we see from Figures 13-16 the point of intersection corresponding to
this mass has $j_{\theta}\approx 0$ for all $\Theta_{\infty}$.

\section{Discussion and conclusions}

We have obtained the cross sections of the amount of mass lost by a
star tidally disrupted by a rotating black hole in the range of black
hole masses where the effects of General Relativity are
important. Our results can be used in a situation where the process
of direct capture of the stars by the black hole and the process of
tidal disruption compete with each other.    
With a high accuracy the tidal cross sections in the plane
of angular momenta $(j_{\theta}, j_{\phi})$ are circles
shifted toward negative values of $j_{\theta}$ for $a\ne 0$. 
We have found the values of 
radius and the shift of these circles as functions of the black hole
mass $M$ for several values of the rotational parameter $a$.  
The ratio of characteristic size of the tidal cross section to the
size of the capture cross sections decreases with the mass $M$. 
Since the shift of the capture cross section toward negative values of
$j_{\theta}$ is always larger than the shift of the tidal cross sections, the
cross sections intersect each other at a certain value of the mass. 
The intersection point moves toward positive values of
$j_{\theta}$ and when the black hole mass is sufficiently large: 
$M > M_{+}(a)$, the process of tidal disruption is no longer possible. 
We estimate $M_{+}(a=0) \sim 4\cdot 10^{7}M_{\odot}$ and 
$M_{+}(a=1) \sim 10^{9}M_{\odot}$. 

We have used the numerical model of a tidally disrupted star developed in
our previous work (IN, ICN). Approximately $5\cdot 10^{3}$ tidal
encounters have been considered, the
number at least an order of magnitude larger than all results obtained 
elsewhere. 

Our results would be helpful
for detailed studies of tidal feeding of the central
black holes by the stellar gas. 
  
To model the star we use the simple $n=1.5$ polytrope. Since the stars
are also often approximated by polytropes with larger values of $n$,
it will be interesting to obtain the dependence of the results on
$n$. Such a study lies beyond the scope of the present Paper. We note,
however, that the polytropes with larger $n$ are more centrally
condensed, and therefore, it is more difficult to disrupt them.
Accordingly, the mass $M_{+}(a)$ should be smaller for larger $n$.

We consider the stars having  solar mass $M_{\odot}$ and 
solar radius $R_{\odot}$. However, our results 
can be used for  stars with different masses $m$ and radii
$R_{st}$. Indeed, the stellar mass and the
radius enter in our calculations only 
in combination $t_{st}/t_{gr}$, where
$t_{st}=\sqrt {{R_{st}^{3}\over Gm}}$, and $t_{gr}={GM\over c^{3}}$.
Therefore, when the stellar mass and radius are not the solar ones,
the cross sections obtained for the mass $M$ correspond to the
re-scaled mass of the black hole
\be M_{re}=\sqrt{{M_{\odot}\over m}}{({R_{st}\over R_{\odot}})}^{3/2}M.
\label{ieq 1}\ee

As we have shown there is a rather significant region in the plane  
 $(j_{\theta}, j_{\phi})$ where only a partial stripping of mass from
the star takes place. The future of a part of
the star that remains gravitationally bound
after the tidal stripping depends on many factors. Let us suppose that
the orbit of this remainder is unchanged after the tidal encounter. 
Taking into account
that the central density of the remainder is smaller than the density
of the unperturbed star and the remainder has significant internal
motions and rotation, one can suppose that the remainder would be tidally
disrupted during the next tidal encounter. However, the processes of
distant gravitational interactions with other stars of the central 
stellar cluster could increase the value of the orbital angular
momentum of the remainder and it could be present in the cluster for a
certain period of time. The observational detection of such remainders
e. g. in the centre of our own Galaxy would be a very convincing test of
the theory of tidal disruptions of stars by a black hole.

Quantitative results obtained above
should be taken with  caution. From the results of comparison between 
our 
'old' variant of the  model of a tidally disrupted star formulated and 
developed in IN and our 'new' advanced variant formulated and developed in ICN 
it follows that when the
amount of mass lost by the star 
is large enough the 'old' model gives a larger value of the mass
loss, and the difference 
between these models is of the order of $30\%$. Also, the
difference between the 3D hydro-simulations themselves can be as large
as $\sim 50\%$ or even larger 
depending on  particular numerical scheme and 
criterion for the mass loss (e.g Diener et al. 1997, and the
discussion above). However, we believe that our qualitative
conclusions are robust, and  future work on improvement of
the numerical schemes  would mainly lead to 
corrections of the numerical values characterising the process of
the tidal disruption, such as e. g. the value of $M_{+}(a)$.  

We neglect the possibility of
formation of strong shocks in the star. Therefore, our results may 
be changed if future finite difference simulations  will show that 
 shocks play a significant role in the range of parameters we consider.

We have calculated only one quantity related to the outcome of the tidal
disruption event: the amount of mass lost by the star, 
$M_{lost}$. As has been pointed out by
Lacy et al. (1982) and Rees (1988) the dynamics of the gas lost by the
star depends significantly on the distribution of this gas over the
orbital energies. The gas elements with negative orbital energies 
form an eccentric disc and can, in principle, accrete onto the black
hole
\footnote{Note that the fraction of stellar gas eventually accreted
onto the black hole may differ from the fraction of gravitationally 
bound gas lost by the star. The physical processes occurring in the gas
after the tidal disruption event have been discussed by e. g. Evans
$\&$ Kochanek (1989).}. 
The gas elements with positive orbital energies may leave the
gravitational field of the black hole. Therefore, in a future work
we will be calculate the cross sections characterising the 
distribution of the gas lost by the star over the orbital integrals of
motion. Also, it
would be interesting to calculate the cross sections for 
other quantities, such as the energy, internal angular momentum
and average density of the gravitationally bound remainder.

\section*{Acknowledgements}
PBI thanks the INTEGRAL Science Data Centre for hospitality. This work has been supported in part
by RFBR grant 04-02-17444. We are grateful to Alexei Ulyanov and Simon Shaw for
useful remarks.

\end{document}